\documentclass[11pt]{article}

\def\A{\mathsf{A}}
\def\B{\mathsf{B}}
\def\F{\mathsf{F}}
\def\K{\mathsf{K}}
\def\J{\mathsf{J}}
\def\P{\mathsf{P}}
\def\L{\mathsf{L}}
\def\SO{\mathsf{SO}}
\def\so{\mathsf{so}}
\def\g{\mathsf{g}}

\def\h{\mathsf{h}}

\def\v{\mathsf{v}}

\def\l{\ell}
\def\cD{{\cal D}}

\def\f{\frac}

\def\extd{\mathrm{d}}

\def\tr{\mbox{Tr}}

\begin{document}

\title{%
Can we see gravitational collapse in (quantum) gravity perturbation
theory?}
\author{%
 J.\ Kowalski--Glikman\thanks{Institute for Theoretical Physics,
University of Wroclaw,   Poland;  {\tt
jkowalskiglikman@ift.uni.wroc.pl}},%
 $\;$ and A.\ Starodubtsev\thanks{Institute for Theoretical
 Physics,
University of Utrecht, The Netherlands;  {\tt
A.Starodubtsev@phys.uu.nl}}%
} \maketitle

\begin{abstract}
In this paper, by making use of the perturbative expansion around
topological field theory we are trying to understand why the
standard perturbation theory for General Relativity, which starts
with linearized gravity does not see gravitational collapse. We
start with investigating classical equations of motion. For zero
Immirzi parameter the ambiguity of the standard perturbative
expansion is reproduced. This ambiguity is related to the appearance
of the linearized diffeomorphism symmetry, which becomes unlinked
from the original diffeomorphism symmetry. Introducing Immirzi
parameter makes it possible to restore the link between these two
symmetries and thus removes the ambiguity, but at the cost of making
classical perturbation theory rather intractable. Then we argue that
the two main sources of complexity of perturbation theory, infinite
number of degrees of freedom and non-trivial curvature of the phase
space of General Relativity could be disentangled when studying {\it
quantum} amplitudes. As an illustration we consider zero order
approximation in quantum perturbation theory. We identify relevant
observables, and sketch their quantization. We find some indications
that this zero order approximation might be described by Doubly
Special Relativity.
\end{abstract}

\newpage

\section{Introduction}

One of the simplest models describing  a collapse of a spacetime is
$2+1$ dimensional gravity coupled to point particles \cite{2+1grav}.
A particle induces a conical singularity with the deficit angle
proportional to its mass. If cosmological constant is zero spacetime
collapses when deficit angle becomes larger than $2\pi$. Physically
it results in a bound on energy, or in the appearance of an
invariant energy scale in the relativistic symmetries of particles
propagation in such spacetime. The latter are known as Doubly
Special Relativity symmetry \cite{giovanni} (see also \cite{dsr}.)

The picture of collapsing spacetime becomes more elaborate if we
add cosmological constant. If $\Lambda > 0$ spacetime has an outer
cosmological horizon. When we add a massive particle the
cosmological horizon shrinks around it, and if the mass of the
particle exceeds Planckian the horizon shrinks to a point. If
$\Lambda < 0$, the energy is not bounded from above. However as
soon as mass of a particle gets larger than $m_{Pl}$, it becomes
surrounded by a horizon known as the BTZ black hole.

In four spacetime dimensions the picture of gravitational collapse
is even more intricate. Classically, any point particles creates a
black hole around it with its size proportional to its mass. In
the presence of positive cosmological constant one has also a
cosmological horizon which shrinks  as the mass grows. One can
even imagine a situation when the horizon of a black hole collides
with cosmological horizon. This could be a bound on energy in four
dimensional GR analogous to the energy bound in 2+1 gravity
mentioned above.

It is worth mentioning here that as it is understood at least in
$2+1$ gravity the phenomena described above are vital for
consistent quantization of the theory. The energy bound is
responsible for the appearance of fixed point in renormalization
group.\footnote{Sometimes it is said that $2+1$ gravity owes  its
renormalizibility to the absence of local degrees of freedom. This
is not the case however. A simple counterexample is a quantization
of $2+1$ gravity by expanding around a fixed background. It knows
about the absence of propagating modes of gravitational field, but
it is still non-renormalizible}

The problem is that the standard perturbative approach which starts
by linearizing gravity around a fixed background can see none of
these phenomena. If in $2+1$ dimensions one can hope to solve the
theory exactly, this seems hardly possible in dimension four. There
the complexity of the theory is multiplied by the presence of
infinite number of degrees of freedom. The natural question then
arises: is it possible to develop an approximate method to solving
(quantum) Einstein's equations which would be able to see
gravitational collapse, maybe in some simplified form? The first
step towards answering this question would be to understand the
reason why the standard perturbative approach does not manage to do
it.

We try to address this question in the present paper. For this we
utilize the perturbative framework of \cite{Freidel:2005ak} which
uses a certain topological field theory (namely BF theory) as a
starting point. It requires  a non-zero cosmological constant and
also admits inclusion of Immirzi parameter.

We start by studying such perturbative expansion of classical
equations of motion with zero Immirzi parameter. It appears that
it reproduces the results of standard perturbation theory which
starts with linearization. The linearized Einstein's equations
emerge in it as certain integrability conditions for the first
order equations.

The immediate problem that appears at this stage is the invariance
of this equations with respect to linearized diffeomorphisms which
become unlinked from the actual diffeomorphisms and therefore do
not correspond to any physical symmetry. This leads to ambiguities
in the physical results obtained from such perturbation theory.

This may seem strange for an approach starting from BF theory as
it is invariant with respect to exact diffeomorphisms. However
linearized diffeomorphism transformation now appear in a different
guise: as a part of "translational" symmetry group of BF-theory
which is dual to the local gauge symmetry.

A small modification of BF theory, namely introducing Immirzi
parameter provides a natural solution to this problem. It connects
the "translational" symmetry and the local gauge symmetry into a
single one, and thus, to first approximation, reestablishes the
link between linearized and actual diffeomorphisms. It has a
physical effect that stringy degrees of freedom coming from
breaking "translational" symmetry now become coupled to particles
degrees of freedom coming from breaking local gauge symmetry.

 However the next
problem emerges here. The first order integrability conditions turn
into the form of exact Einstein's equations and as such cannot be
solved in a generic situations. This makes the classical
perturbation theory not useful in practice.

Fortunately the situation is very different in quantum theory where
we are interested not in calculating the values of the fields
everywhere in space, but only in scattering amplitudes for finite
number of particles. At a finite order of perturbation theory only
finite number of modes of the field contribute to such amplitudes.
This allows us to reduce the problem to finite-dimensional, while
keeping track of some non-linearities of General Relativity at the
same time.

One of the effects of introducing Immirzi parameter in this case is
that quantum perturbation theory becomes ``semi-holographical''. Its
Feynman diagrams map on a boundary of space and they braid as in 3
dimensions. At the same time every particle in a diagram has 3
degrees of freedom as in 4 dimensions. In Euclidian theory for
example  the braiding effect makes energy and momenta bounded from
above. The bound is controlled by a certain combination of Newton
constant, cosmological constant, and Immirzi parameter.

In section 2 we review the setup of \cite{Freidel:2005ak} and
\cite{Freidel:2006hv} for perturbation theory for gravity and its
coupling to particles.

In section 3 we solve it to the first order for zero Immirzi
parameter and identify the problem related to the appearance of an
extra gauge symmetry.

In section 4 we discuss physical implication of this extra gauge
symmetry, which is decoupling of some of the degrees of freedom
linked in the full theory, and show how introducing Immirzi
parameter restores this coupling. We also show that Immirzi
parameter mixes mass with Taub-NUT charge, which results in the
appearance of non-local singularities -- gravitational Dirac
strings.

In section 5 we give a brief perspective on how to calculate
particle scattering amplitudes in this perturbation theory. We
identify the observables which contribute to perturbative
calculation of the amplitudes and show that at any finite order
there is only a finite number of them, which makes quantum
perturbation theory tractable. For non-zero Immirzi parameter this
observables become non-local due to Dirac strings, and we argue that
this leads to a certain kind of "dimensional reduction" in quantum
amplitudes. Finally, we take a closer look at the observables which
are relevant at zero order of perturbation theory. We show that the
variables canonically conjugate to particle positions form  de
Sitter space, which is an indication of the presence of Doubly
Special Relativity effects in the theory.

In section 6 we discuss some mathematical problems which has to be
solved to complete the definition of these perturbation theory.

\section{Equations for gravity with particle source}

To set the stage, let us recall, following \cite{Freidel:2005ak} and
\cite{Freidel:2006hv} the formulation of gravity as a constrained
topological field theory and it coupling to point particle(s). Let
us start with the pure gravitational field, which is encoded in a
$\SO(4,1)$ (de Sitter) connection
\begin{equation}\label{1}
 \A_\mu=A^{IJ}{T_{IJ}}=
\left(\f{1}{\l}e_\mu{}^a\,T_{a4} +\f{1}{4} \omega_\mu{}^{ab}{}\,
T_{ab}\right)
\end{equation}
where index $a$ runs from $0$ to $3$. In this formula $T_{IJ}$, $I,J
= 0,\ldots 4$ are generators  of  Lie algebra $\so(4,1)$.
$e_{\mu}{}^{a}$ is the frame field from which the metric is
constructed and $\omega_{\mu}{}^{ab}$ is the spin connection.
Connection has a natural mass dimension 1 whereas the frame field is
dimensionless; this is the reason why a length scale $\l$ appears in
the expression of the components of $\A$ representing the frame
field. This length scale has to be the cosmological length $\l$
related to the cosmological constant by $$ \frac{1}{\l^2}=
\frac{\Lambda}{3} $$

To formulate the theory  we also need a two form valued in the Lie
algebra $\so(4,1)$ denoted by $\B = B_{\mu \nu}{}^{IJ}{T_{IJ}}\,
\extd x^\mu\wedge \extd x^\nu$. In terms of this two form and
connection $\A$ the action takes the form
\begin{equation}\label{2}
S=\int \left(B_{IJ}\wedge F^{IJ} -\frac{\alpha}{4}\,
\epsilon^{IJKLM}\, B_{IJ}\wedge B_{KL}\, v_M-  \frac{\beta}{2}\,
B^{IJ}\wedge B_{IJ}\right) \label{2.1a}
\end{equation}
with $v_M$ being a constant vector in fundamental representation
of the gauge group, forcing the gauge symmetry breaking from
$\SO(4,1)$ down to $\SO(3,1)$ (see \cite{Smolin:2003qu}.) This
action can be shown \cite{Freidel:2005ak} to be equivalent to the
action of General Relativity with nonzero cosmological constant
and a nonzero, dimensionless Immirzi parameter
 $\gamma$.
The initial parameters dimensionless $\alpha, \beta, \l$ are related
to the physical ones as follows
\begin{equation}\label{3}
\frac{1}{\l^2}=\frac{\Lambda}{3},\qquad \alpha =
\frac{G\Lambda}{3}\frac{1}{(1+\gamma^2)}, \qquad \beta =
\frac{G\Lambda}{3}\frac{\gamma}{(1+\gamma^2)}
\end{equation}

Let us now turn to the particle(s) coupling. As it was shown in
\cite{Freidel:2006hv} matter can arise in the most natural way  by
introducing the simplest possible term breaking the gauge symmetry
of the theory in a localized way. The gauge degrees of freedom are
then promoted to dynamical degree of freedom, which reproduce the
dynamics of a relativistic particle coupled to gravity. This idea is
realized is obtained by choosing a worldline $P$ and a fixed element
$\K$ of the $\so(4,1)$ Lie algebra with generators $T^{IJ}$,
depending on the particle rest mass and spin
\begin{equation}\label{4}
    \K\equiv m \l\, T^{04} + s
T^{23}
\end{equation}
so as to have
\begin{equation}\label{5}
    S_{P}(\A) =-\int
\extd \tau\, \tr \left(\K \A_\tau(\tau)\right)
\end{equation}
where $\tau$ parameterizes the world line $z^\mu(\tau)$ and
$\A_\tau(\tau) \equiv \A_\mu(z(\tau))\, \dot{z}^\mu$.

The action for the particle is obtained from (\ref{5}) by adding
gauge degrees of freedom that become dynamical at the particle
worldline. To this end we take $\A^\h=\h^{-1} \A \h + \h^{-1} d \h$
to be the gauge  transformation of $\A$. Then the particle
lagrangian takes the simple form
\begin{equation}\label{6}
    L(z,\h; \A) = -\tr
\left(\K \A^\h_\tau(\tau)\right)\quad S = \int\, d\tau\, L(z,\h; \A)
\end{equation}
which can be rewritten as
\begin{equation}
L(z,\h; \A)=-{\tr}(\J \A_\tau) + L_1(z,\h) \label{7}
\end{equation}
where in the first term  $\J$ is given by
\begin{equation}\label{8}
    \J \equiv
\h\, {\K} \,\h^{-1}
\end{equation}
 The first term in equation (\ref{7}) describes the
covariant coupling between the particle and the $\A$ connection of
the (constrained) BF theory, while the second
\begin{equation}
L_1(z,\h) =  -\tr(\h^{-1}\dot \h \K), \label{9}
\end{equation}
describes the dynamics of the particle.

The field equations for gravity coupled to point particle, resulting
from (\ref{2}) and (\ref{7}) in the gauge, in which the particle is
at rest at the origin of the coordinate system take the form
\begin{equation}\label{10}
    F^{IJ} = \alpha\, \epsilon^{IJKLM} \, B_{JK}\, v_M + \beta
    B^{IJ}
\end{equation}
\begin{equation}\label{11}
   \cD_\A\, B^{IJ} = J^{IJ}\, \delta_P, \quad \delta_P =\delta^3(x) \varepsilon
\end{equation}
where $\cD_\A$ is covariant derivative of connection $\A$ and
$\varepsilon$ is the volume three-form on constant time surface.

In this paper we will be primarily interested in solving these
equations in the limit $\alpha\rightarrow0$. It would seem that it
suffices to put $\alpha=0$ in (\ref{10}) and solve the resulting
equations; however there could be contributions from higher order
equations affecting this limit. Therefore more care is needed and
we consider expansion of the equations up to the first order.

Consider the expansion in $\alpha$
$$
\A = \A^{(0)} + \alpha \A^{(1)} + \ldots, \quad \B = \B^{(0)} +
\alpha \B^{(1)} + \ldots
$$
In the zero order in $\alpha$ we get
\begin{equation}\label{12}
    F^{IJ}(\A^{(0)}) =  \beta
    B^{(0)}{}^{IJ}
\end{equation}
\begin{equation}\label{13}
   \cD_{\A^{(0)}}\, B^{(0)}{}^{IJ} = J^{IJ}
\end{equation}
In the first order in $\alpha$ we find
\begin{equation}\label{14}
  \cD_{\A^{(0)}}\,  \A^{(1)} = \epsilon^{IJKLM} \, B^{(0)}{}_{JK}\, v_M  + \beta
    B^{(1)}{}^{IJ}
\end{equation}
\begin{equation}\label{15}
   \cD_{\A^{(0)}}\, B^{(1)}{}^{IJ}+ [\A^{(1)},B^{(0)}]^{IJ} = 0
\end{equation}
If we now take the covariant differential $\cD_{\A^{(0)}}$ of
(\ref{14}) and compare it with (\ref{15}), we find the first
integrability condition
\begin{equation}\label{16}
\epsilon^{IJKLM} \,\cD_{\A^{(0)}}\, (B^{(0)}{}_{JK}\, v_M) =0.
\end{equation}
The second integrability condition comes from acting by
$\cD_{\A^{(0)}}$ on eq.(\ref{15}) and reads
\begin{equation}\label{16a}
[\B^{(0)},\B^{(0)*}]^{IJ}+[\A^{(1)},\J]^{IJ}=0,
\end{equation}
where
\begin{equation}
B^{(0)*IJ}=\epsilon^{IJKLM}B^{(0)}{}_{JK}\, v_M
\end{equation}

Equations (\ref{16}), (\ref{16a}) affect the zero order
contributions to the field $\B^{(0)}$ and cannot be ignored even
in the limit $\alpha\to 0$.

 Thus the topological limit of gravity, coupled to a
point particle is governed by four equations (\ref{12}), (\ref{13}),
(\ref{16}), (\ref{16a}). In what follows we will drop the ${}^{(0)}$
superscript to simplify the notation.

\section{Zero Immirzi parameter and ambiguities in the solution}
In this section we solve the above equations to the first order
for $\beta=0$ and identify the problem arising in such
approximation.

Zero order is described by equations (\ref{12}),(\ref{13}). For
$\beta=0$ those equations possess the following gauge symmetry:
\begin{equation}\label{gaugesym0a}
\A \to \g^{-1}\A \g+\g^{-1}d \g, \ \ \ \B \to \g^{-1}\B\g
\end{equation}
and
\begin{equation}\label{gaugesym0b}
 \B \to \B + \cD_\A \Phi
\end{equation}
where $g\in \SO(4,1)$, and $\Phi$ is a $\so(4,1)$-valued one form.

The solution in the zeroth order is thus
\begin{equation}\label{solution0}
\A =\g^{-1}d \g \ \ \ \ \B = \B_p + \g^{-1} \Phi \g,
\end{equation}
where $\B_p$ is a particular solution depending on $\J$. We
consider spinless particle, so we choose $\J = \g^{-1} K\, g =
m\ell \g^{-1} T^{04}\,\g$.

If we define  $\bar \B_p =\g \B_p \g^{-1}$ the above equation
simplifies to
\begin{equation}\label{21}
    d\bar \B_p=\K
\end{equation}
and a possible solution is
\begin{equation}\label{23}
  \bar \B_p = m\ell\, T^{04}\, \epsilon_{abc}\, \frac{x^a}{r^3}\,
   dx^b\wedge dx^c
\end{equation}

So far $\g$ and $\Phi$ were arbitrary, but from the first order we
have extra equations (\ref{16}), (\ref{16a}). The origin of this
equations can also be explained the following way. We take the
general solution of zeroth order equation (\ref{solution0}) and
plug it into the symmetry breaking term in (\ref{2}) (the one
which is proportional to $\alpha$). Then by varying the resulting
expression w.r.t. $\g$ we obtain equation (\ref{16a}) and by
varying it w.r.t. $\Phi$ we obtain (\ref{16}).

However because the whole action is diffeomorphism invariant and
because the translational part of $\g$ is closely related to
diffeomorphism transformations we can start by making an arbitrary
choice of $\g$ which will simply result in a choice of coordinate
system.

For spherical static coordinates in deSitter space we have to
choose
\begin{equation}\label{20}
\g = \left(
      \begin{array}{ccccl}
        \cosh\frac t\ell & \frac r\ell \sinh\frac t\ell& 0 & 0 & \sqrt{1-\frac{ r^2}{\ell^2}} \sinh\frac t\ell\\
        0 & \sqrt{1-\frac{ r^2}{\ell^2}} \sin\theta\cos\phi & -\sin\phi & -\cos\theta\cos\phi  & -\frac r\ell\sin\theta\cos\phi  \\
        0 & \sqrt{1-\frac{ r^2}{\ell^2}} \sin\theta\sin\phi & \cos\phi & -\cos\theta \sin\phi& -\frac r\ell\sin\theta\sin\phi \\
        0 & \sqrt{1-\frac{ r^2}{\ell^2}} \cos\theta & 0 & \sin\theta & -\frac r\ell \cos\theta\\
        \sinh\frac t\ell & \frac r\ell \cosh\frac t\ell& 0 & 0 &\sqrt{1-\frac{ r^2}{\ell^2}} \cosh\frac t\ell \\
      \end{array}
    \right)
\end{equation}

It remains now to plug the solution (\ref{23}) to first
integrability condition (\ref{16}) and solve it for $\Phi$. For
$\beta = 0$ $\cD_{\bar\A}$ in (\ref{16}) becomes ordinary
differential $d$, and the equation reminds the linearized Einstein
equation, but with unusual type of source. In this case the solution
for $\Phi$ reads
$$
   \Phi =-\ell\left(1-\frac{
   r^2}{\ell^2}\right)\epsilon_{ijb}\frac{x^b}{r^3}\, dt \, T^{ij}
$$\begin{equation}\label{24}
   -\ell\sqrt{1-\frac{
   r^2}{\ell^2}}\,\epsilon_{iab}\frac{x^b}{r^3}\, dx^a \left(\sinh\frac
   t\ell\, T^{4i} +\cosh\frac
   t\ell\, T^{0i}\right)
\end{equation}
Notice that to this point the presence of non-zero cosmological
constant was required. For $\Lambda \to 0$ one has $l\to \infty$
and the expression  (\ref{24}) for $\Phi$  diverges.

Having solved the condition (\ref{16}) one can turn to finding
$\A^{(1)}$ from (\ref{14}). The later equation has a regular limit
as $l\to \infty$ and we can take it at this point. We will write
down explicitly only the metrical part of it, i.e. the part
proportional to $T^{4a}$

\begin{equation}\label{newton}
\A^{(1)}=-\frac{1}{r}T^{40}dt+\frac{1}{r}T^{4i}dx^i+(...)T^{ab}+\cD_{\A^{(0)}}\Psi,
\end{equation}
where $\Psi$ is an arbitrary 0-form field. This expression
contains the standard Newtonian potential term, but there is also
some ambiguity in it. Different choices of $\Psi$ lead to
different metric that cannot be related by coordinate
transformation. This is a source of ambiguity in physical results.

Similar problem can be observed in linearized gravity as well.
Linearized Einstein's equations
\begin{equation}\label{linearized}
\nabla^m \nabla_m
h_{ab}-2\nabla^m\nabla_{(a}h_{b)m}+\nabla_a\nabla_b
(g^{mn}h_{mn})=0
\end{equation}
are invariant with respect to exact diffeomorphisms when one
simultaneously transforms the background field $g_0^{ab}\to
g_0^{mn}\frac{\partial x'^a}{\partial x_m}\frac{\partial
x'^b}{\partial x_n}$ and the fluctuation $h_{ab}\to
h_{mn}\frac{\partial x^m}{\partial x'_a}\frac{\partial x^n}{\partial
x'_b}$. At the same time they are invariant with respect to
linearized diffeomorphisms when one transforms fluctuations only
$h_{ab}\to h_{ab}+\nabla_{(a}\xi_{b)}$, where $\xi_a$ is an
increment of a coordinate $x_a$, keeping the background field fixed.
Linearized gravity doesn't "remember" that the two above symmetries
have common origin. Therefore the gauge-fixing for both symmetries
can be done independently.

To illustrate one of the consequences of the above ambiguity let
us recall one curious fact that the Schwarzschild solution of the
Einstein's equations in Kerr-Schild parameterization is also a
solution of linearized Einstein's equations
(\ref{linearized})\cite{ksx}. It is related to the usual
Schwarzschild coordinates by transformation $T=t-2m\ln(r/2m-1)$,
and the metric in this parameterization looks like
\begin{equation}
ds^2=-\Big(1-\frac{2m}{r}
\Big)dT^2-\frac{4m}{r}dTdr+\Big(1+\frac{2m}{r} \Big)dr^2 +r^2
d\Omega.
\end{equation}
It is a solution of  equations (\ref{linearized}), and it looks
quite similar to the standard Newtonian metric
\begin{equation}
ds^2=-\Big(1-\frac{2m}{r} \Big)dT^2+\Big(1+\frac{2m}{r} \Big)dr^2
+r^2 d\Omega,
\end{equation}
which is also a solution of (\ref{linearized}). The two however
cannot be related by a coordinate transformation, and they are
physically different with different Penrose diagrams. They can be
related by linearized diffeomorphism transformation with
$\xi_0=-4m\ln(r/4m)$ and $\xi_i=0$.

Clearly the arbitrariness in the choice of $\Psi$ in
(\ref{newton}) leads to exactly the same ambiguity as above. By
adjusting $\Psi$ we can have either standard newtonian potential
or a black hole. This is a problem because perturbation theory
cannot distinguish between the two.

However we also have to  remember about the second integrability
condition (\ref{16a}). By direct calculation one can check that away
from particle sources it is automatically satisfied. This was to be
expected as a result of diffeomorphism invariance. At the points of
the particle sources this condition reduces to
\begin{equation}\label{int2out}
[\A^{(1)}+\Phi^*,\J]^{IJ}=0,
\end{equation}
where
\begin{equation}
\Phi^{*IJ}=\epsilon^{IJKLM}\Phi_{JK}\, v_M
\end{equation}

As it was shown in \cite{Freidel:2006hv} equation (\ref{int2out})
is directly related to geodesic equation for the test particles.

One can see that in equation (\ref{int2out}) the problem of
invariance of equation (\ref{14}) with respect to linearized gauge
transformation $\A^{(1)}\to \A^{(1)}+\cD_A\Psi$ is combined with
another ambiguity which is the invariance of equation (\ref{16})
for $\Psi$ with respect to similar transformation
\begin{equation}\label{gchi}
\Phi\to\Phi +\cD_A\chi,
\end{equation}
 where $\chi$ is an arbitrary scalar field.
But the later symmetry is a subgroup of the symmetry of $BF$-theory,
namely  (\ref{gaugesym0b}). Thus, the symmetry that causes the
problem in the linearized gravity has its counterpart in BF theory.
Therefore, the perturbative approach starting with BF theory
considered in this section will suffer from the same problem as the
standard perturbation theory that starts from linearized gravity.
Similar conclusion has been reached also in \cite{rovelli1} in the
case of an analogous treatment of Yang-Mills theory.

However the considerations of this section provide a clear hint on
how this problem is to be cured within  perturbative framework. This
is the subject of the next section.

\section{Non-zero Immirzi parameter  and restoration of unique gauge symmetry}

Let us first discuss the physical meaning of the gauge symmetries
that appear in BF theory. Initially the gauge parameters physically
irrelevant as such, of course. However, after introducing matter by
applying the symmetry breaking perturbation some of the gauge
parameters turn into physical degrees of freedom. The easiest way to
look at them is to introduce localized symmetry breaking terms,
which results in the appearance of finite number of degrees of
freedom, similar to those we used introducing particles in section
1. Then one can ask: what is the physical meaning of the degrees of
freedom resulting from such symmetry breaking.

Consider $BF$ theory coupled to a most general type of localized
sources, which includes particles considered above as well as
strings considered in \cite{bfstrings}.

\begin{equation}\label{bfwsources}
S=\int B_{IJ}\wedge F^{IJ}+S_P(\A)+S_S(\B),
\end{equation}
where $S_P(\A)$ is the same as in (\ref{5}), and the string action
reads
\begin{equation}\label{stringaction}
S_S(\B)=\int d\tau d\sigma \tr (\L \B_{\tau \sigma}),
\end{equation}
where $\tau$ and $\sigma$ are variables parameterizing the string
worldsheet, $\B_{\tau \sigma}\equiv \B_{\mu \nu} \frac{\partial
x^\mu }{\partial \tau}\frac{\partial x^\nu }{\partial \sigma}$,
and $\L$ is a charge analogous to (\ref{4}) but so far with
different mass and spin.

Equations of motion for particles were described in section 2. To
obtain equations of motion for stringy source one has to perform a
gauge transformation $\B \to \h^{-1}(\B+d \Phi)\h$ and substitute
it back to the string action (\ref{stringaction}). After
integration by part the string action will look like
\begin{equation}\label{stringaction1}
S_S= \int_{W} d\tau d\sigma \tr (\P \B_{\tau
\sigma})+\int_{\partial W}d\tau \tr(\Phi \P)- \int_{W} d\tau
d\sigma \tr (\Phi d_A \P ),
\end{equation}
where $\P=\h \L \h^{-1}$ is the string momentum density. The
equations following from action (\ref{stringaction1}) have to be
supplied with boundary conditions at the string endpoints. To
allow for non-vanishing string momentum density the natural
condition is
\begin{equation}\label{sbc}
\Phi\Big\vert_{\partial W}=\cD_A \chi. \end{equation} Thus, from
the third term in (\ref{stringaction1}) we get conservation of
string momentum density along the whole worldsheet,
\begin{equation}
d_A \P=0,
\end{equation}
and from the second term in (\ref{stringaction1}) we get the same
condition, but restricted to string endpoints. The variational
principle thus become consistent for arbitrary string momentum $\P$.
From what the second term in (\ref{stringaction1}) with boundary
condition (\ref{sbc}) turns into one can deduce that the gauge
parameter $\chi$ is canonically conjugate to the string momentum
$\P$ at its endpoints. Notice that the gauge parameter $\chi$ can be
turned into a physical degree of freedom only at the endpoint of the
string. Thus, it becomes somewhat analogous to Chan-Patton degrees
of freedom in string theory.

So far the particle positions $\h$ and  momenta $\J$ are completely
independent from string momenta $\P$ and their canonical conjugates
$\chi$. This is a consequence of a split between exact gauge
transformations encoded in $\h$ and linearized gauge transformations
encoded in $\chi$. Below we consider how the link between the two
can be restored already at zero order of perturbation theory.

Consider the constraints of the theory described by action
(\ref{bfwsources}). These are equations (\ref{12}), (\ref{13})
reduced on the spacial slice $\Sigma$ with sources on the r.h.s.
(so far we are considering $\beta=0$ case).
\begin{equation}\label{12c}
    F^{IJ}(\A) =P^{IJ}\delta_S
\end{equation}
where $\delta_S=\int_{W\cap\Sigma}d\sigma \delta^3(x-x(\sigma))$,
and
\begin{equation}\label{13c}
   \cD_{\A}\, B^{IJ} = J^{IJ}\delta_P,
\end{equation}
where $J^{IJ}$ is a particle source described in section 2.
Constraint (\ref{13c}) generates gauge transformations
(\ref{gaugesym0a}), and constraint (\ref{12c}) -- transformations
(\ref{gaugesym0b}). The later contains a subgroup (\ref{gchi}),
which is directly linearized diffeomorphism transformations that
were detected in the previous section as a source of the problem.

The obvious approach to solving this problem is to make the theory
remember that the two symmetries have common origin, i.e. that the
constraints  (\ref{12c}) and (\ref{13c}) generating them are not
actually independent. This is what naturally happens when we
introduce a $\beta$ term in (\ref{12c})
\begin{equation}\label{12c1}
    F^{IJ}(\A)+\beta B^{IJ} =P^{IJ}\delta_S
\end{equation}
(\ref{13c}) becomes a consequence of (\ref{12c1}) due to Bianchi
identity, provided that particle sources are located at the
endpoints of the strings. The string becomes something like the
Dirac string for "magnetic" charge $\J$. The role of the Dirac
strings in this context we will discuss below in detail. Positions
of the particles and positions of the strings are thus no longer
independent. At the same time the gauge transformations
(\ref{gaugesym0b}) get deformed and also start affecting connection.
The constraint (\ref{12c1}) generates the following transformations
\begin{equation}\label{gaugesym0bm}
 \B \to \B + \cD_\A \Phi+ \beta [\Phi,\Phi], \ \ \ \A \to \A+\beta
 \Phi,
\end{equation}
and the analog of the subgroup (\ref{gchi}) is now
\begin{equation}
\A+\beta \Phi \to \g^{-1}d\g+\g^{-1}(\A+\beta \Phi)\g,
\end{equation}
which is the same as the gauge transformation (\ref{gaugesym0a}).
The link between the original gauge transformation and the
linearized gauge transformation is thus restored. One of the
resulting effects is that particles degrees of freedom and
Chan-Patton degrees of freedom at the strings endpoints have
merged into one.

Let us now take a closer look at  equations (\ref{12}),
(\ref{13}), and (\ref{14}) for $\beta\not=0$. The first, rather
obvious thing to notice is that, as a result of Bianchi identity,
eqs.\ (\ref{12}) and (\ref{13}) cannot be solved for non-trivial
source if the connection is nonsingular. This problem can be
solved by recalling the Dirac magnetic monopole solution (see, for
example, \cite{nakahara}): we assume that our connection contains
the part proportional to Dirac monopole connection
\begin{equation}\label{17}
   A = \g^{-1}d\g +\beta \g^{-1}( A_D + \Phi)\g, \quad A_D \sim a_D \equiv
   (1-\cos\theta)d\phi
\end{equation}
where $g\in \SO(4,1)$, and $\Phi$ is a $\so(4,1)$-valued one form.
We will use the ansatz (\ref{17}) below. In fact the Dirac string
and the string described in the beginning of this section are the
same.

Consider now eq.\ (\ref{16}), which is an integrability condition
for the first order equations. Computing covariant differential, and
noticing that since $v_M$ has only one non-zero component $v_4=1$ we
find
$$
\cD_\A\, v_M = \A_M{}^4= e_i, \quad i=0,\ldots 4
$$
where $e_i$ is the tetrad one form. Now it follows from eqs.\
(\ref{16}) and (\ref{12}), (\ref{13}) that
$$
\frac1\beta\, \epsilon^{IJKLi} \, F_{JK}\wedge e_i +
\epsilon^{IJKLM} \, v_M J_{KL}=0
$$
It can be shown  that for one particle the second term in the
expression above vanishes, and thus we are left with
\begin{equation}\label{18}
\epsilon^{IJKLi} \, F_{JK}\wedge e_i =0
\end{equation}
Decomposing $\F$ into torsion $F_{4i}\equiv T_i$ and curvature
$F_{ij}\equiv R_{ij}$ two-forms one finds that any connection
being a solution of our problem must be torsion-free $T_i =0$ and
satisfy the vacuum Einstein--de Sitter equations\footnote{That is,
Einstein equations with a positive cosmological constant.}
\begin{equation}\label{19}
\epsilon^{ijkl} \, R_{jk}\wedge e_l =0
\end{equation}

Having obtained the exact Einstein's equations  already at the first
order of perturbation theory, one can be sure that non-perturbative
effects, including gravitational collapse cannot be missed. But this
is not a solution of any problem, because doing such perturbation
theory is the same as solving General Relativity non-perturbatively.
Thus, in classical theory, the present perturbative approach does
not lead to any simplification. However it can bring about some new
intuition in quantum theory which is the subject of the next
section.

Before finishing this section let us point out some another
consequence of introducing Immirzi parameter. Let us solve
eq.(\ref{19}) for the ansatz (\ref{17}) to the leading order in
$\beta$, which is the same as linear approximation. In this
approximation the solution for $\Phi$ will be (\ref{24}). One can
check that $\g^{-1}\Phi\g$ does not have terms proportional to
$T^{4c}$ so that the contribution of it to the tetrad one-form is
zero. The only contribution to the tetrad (in the leading order in
$\beta$) is therefore (cf.\ (\ref{17})
\begin{equation}\label{25}
    A_{i4} = e_i = \left(\g^{-1}\,d\g + \J a_D\right)_{i4}
\end{equation}
One can then easily find what the metric $g = \eta^{ij}\, e_i
\otimes e_j$ is
\begin{equation}\label{26}
    g = -\left(1-\frac{
   r^2}{\ell^2}\right)\left(dt + N(1-\cos\theta)d\phi\right)^2 +\left(1-\frac{
   r^2}{\ell^2}\right)^{-1}\, dr^2 + r^2(d\theta^2 + \sin^2\theta\,
   d\phi^2).
\end{equation}
This is de Sitter--Taub--NUT metric linearized with the Taub--NUT
charge $N$ and with no mass. If one goes beyond linear
approximation one obtains the exact Taub--NUT solution. This means
that the Immirzi parameter mixes mass with Taub--NUT charge. The
presence of a non-local (Dirac string) singularity has a profound
consequences in quantum theory. This will be discussed in the next
section.

\section{Quantum amplitudes versus classical solutions}

The two main sources of difficulties with four dimensional General
Relativity are infinite number of degrees of freedom and
non-linearity of its equations. One of the consequences of the
latter is a non-trivial curvature of the phase space of the theory.
One of the lessons from 2+1 gravity coupled to matter is that
momentum part of its phase space is a group manifold rather than
linear space and the information about its curvature is essential
for consistent quantization of the theory.

In standard perturbation theory one starts by choosing a background
-- a certain point in the phase space of the theory. One then
proceeds by linearizing the theory around this background, which in
particular implies that one works in a tangent space of the phase
space of the theory instead of the actual phase space. This space
remains linear even after removing gauge degrees of freedom by
taking its quotient by linearized diffeomorphism transformations.

Another property of quantum perturbation theory  is that when we are
interested in scattering of a finite number of particles in a finite
order of perturbation theory we have to calculate Feynman diagrams.
The expressions for Feynman diagrams are finite dimensional
integrals instead of infinite dimensional path integrals. Thus
quantum theory allows us to select a set of questions which could be
answered by taking into account only finite number of modes of
quantum field. This option is not available in classical
perturbation theory where we have to solve infinite-dimensional
field equations in every order. This is a big advantage of quantum
perturbation theory over the classical one.

Now one can notice that the ability of quantum perturbation theory
to reduce the problem of calculating scattering amplitudes to finite
dimensional one, and its inability to see the curvature of the phase
space of the theory are not logically linked which each other.
Therefore one can ask if there is a quantum perturbation theory
which  allows to reduce the problem to a certain finite-dimensional
subspace of the phase space and if that subspace is curved the
perturbation theory still can see it.

Below we show that the quantum analog of the perturbation theory
studied here does precisely this job, and it allows us to see some
non-trivial effect already in zeroth order.

We start by explaining how the perturbative expressions for
quantum amplitudes get reduced to finite-dimensional integrals.

The general perturbative expression for the partition function
coupled to arbitrary finite number of particles (where we neglect
all the interactions except gravitational) after integrating out
$B$-field looks like
\begin{eqnarray}\label{series1}
Z(\{x_{p_i}\},\{x_{p_f}\})=\int \cD A \sum\limits_n
\frac{(i\alpha)^n}{n!}\Big( \int v_A
\epsilon^{ABCDE}F_{BC}(x)\wedge F_{DE}(x) \Big)^n  \nonumber \\
\times\exp [i \int_M F^{IJ} \wedge F_{IJ}+\sum
S_p(x_{p_i},x_{p_f})],
\end{eqnarray}
Where $S_p$ are actions for particles of the type (\ref{5}), and
 $x_{p_i}$ and $x_{p_f}$ are initial and final points of $p$-th
 particle. They are presumably
located on initial and final slice of $M$. The expression
(\ref{series1}) can be rewritten in terms of n-point functions as
\begin{equation}\label{series2}
Z(\{x_{p_i}\},\{x_{p_f}\})=\sum\limits_n \frac{(i\alpha)^n}{n!}
\int dx_1 ... dx_n W(x_1,...,x_n,\{x_{p_i}\},\{x_{p_f}\}),
\end{equation}
where
\begin{eqnarray}\label{n-point}
W(x_1,...,x_n,\{x_{p_i}\},\{x_{p_f}\})=\int \cD A\prod\limits_n
\Big( v_A \epsilon^{ABCDE}F_{BC}(x_n)\wedge F_{DE}(x_n) \Big)
\nonumber
\\ \exp [i\frac{1}{\beta} \int_M F^{IJ} \wedge F_{IJ}+\sum
S_p(x_{p_i},x_{p_f})]
\end{eqnarray}

Now one has to recall one of the key properties of n-point functions
of diffeomorphism invariant theories, i.e.\ when the action and the
measure are diffeomorphism invariant, (see e.g.\ \cite{rovelli2} and
references therein) that they do not actually depend on $x_n$, as
coordinates do not carry any physical information. The only
dependence may come from possible coincidences of $x_n$, but it
yields a measure zero contribution. Of course n-point functions
depend on relative positions of the points $x_n$, but the latter are
encoded not in the coordinates but in the fields.

This means that the integrands in (\ref{series2}) are constants and
therefore the integrals can simply be dropped, i.e. absorbed into
renormalization of $\alpha$. As a consequence, the $n$-order
contribution to the partition function can be evaluated as a path
integral of topological field theory with $n$ localized symmetry
breaking insertions. As it was discussed e.g.\ in
\cite{Freidel:2006hv}: breaking a symmetry of topological field
theory in a localized way releases only a finite number of degrees
of freedom. This is the central observation which allows us to
reduce the procedure of perturbative calculation of quantum
scattering  amplitudes to a finite dimensional problem. Notice that
this would not be possible in non-perturbative calculation as in
such case the integral with a symmetry breaking term would stay in
the exponent and as a result could not be replaced with a local
symmetry breaking insertion.

Now let us try to compose a list of variables which become
relevant for the calculation of the n-point function
(\ref{n-point}). To simplify the job let us first integrate out
the particles degrees of freedom, $g_p$, which is the gauge
parameter evaluated at the location of the particle. It can be
done by replacing particle terms in the action with particle
propagator.
\begin{eqnarray}\label{propagator}
G_p(g(x_{p_i}),g(x_{p_f}),\A_p)=\int \cD g \exp \Big( \int dt \tr
(\A_t \J) +S_p (g)\Big)
\end{eqnarray}
We will not dwell on calculating this expression here. However, as
it was stressed in \cite{Freidel:2006hv}, the propagator
(\ref{propagator}) is a matrix element of the open Wilson line
operator whose representation is defined by the mass and the spin of
the particle. The states, between which the matrix element has to be
taken are $SO(4,1)$ analogs of Wigner functions, with corresponding
representation indices. Alternatively, one can insert Wilson lines
directly. This would correspond to working in momentum
representation, as the matrix indices of the Wilson line are momenta
of the particles. Eq.\ (\ref{n-point}) now reads:
\begin{eqnarray}\label{n-point1}
W(x_1,...,x_n,\{x_{p_i}\},\{x_{p_f}\})=\int \cD A
\prod\limits_p\big( G_p(g(x_{p_i}),g(x_{p_f}))\big)\nonumber
\\ \prod\limits_n
\big( v_A \epsilon^{ABCDE}F_{BC}(x_n)F_{DE}(x_n) \big)  \exp [i
\int_M F^{IJ} \wedge F_{IJ}],
\end{eqnarray}
The dependence of this expression on $\{x_{n}\}$, $\{x_{p_i}\}$,
and $\{x_{p_f}\}$ should be understood as the dependence on the
connection field evaluated at these points.

Now if instead of connection field $A$ we use its holonomies
connecting different point in space, it is clear that due to
$SO(4,1)$ gauge symmetry of the action in the exponent the integrand
in (\ref{n-point1}) cannot depend on holonomies ending anywhere else
but at the points $x_{p_i}$, $x_{p_f}$, and $x_n$. For convenience
however let us choose one common reference point $x_0$ putting it
for example somewhere on the boundary of initial slice, $x_0 \in
\partial M \cap \Sigma_i$. For basis holonomies we choose $g_n$
connecting $x_n$ to $x_0$, $g_{p_i}$ connecting $x_{p_i}$ to $x_0$,
and $g_{p_f}$ connecting $x_{p_f}$ to $x_0$. Any holonomy connecting
any pair of points can be expressed as a combination of those from
the above list. If we took $\beta=0$ limit, the free action would
force the connection to be flat, and the holonomies would become
independent of pathes along which they are taken. That means that
these holonomies would comprise the full list of variables relevant
for calculating the amplitude (\ref{n-point1}).

The situation is very different when we consider $\beta \not= 0$.
First of all as it was discussed in \cite{Freidel:2006hv} the $BF$
theory with nonzero $\beta$ cannot be consistently coupled to a
local symmetry breaking source without introducing introducing an
extended connection singularity -- the Dirac string.

Generally, if in classical theory the equations of motion are
inconsistent because of a charge non-conservation, in quantum
theory in analogous situation one obtains a vanishing partition
function. The situation is similar here. For example, if  a
particle located away from the boundary, and the connection is
non-singular, in the partition function
\begin{equation}\
Z=\int \cD A \exp \Big(i\int F^{IJ}\wedge F_{IJ} + i\tr (\A
\K)\delta^3 (x-z(t)) \Big)
\end{equation}
The first term in the exponent will be independent of $\A(z(t))$,
and the contribution from integrating over it will be of the form
$\delta (\K)$. This means that if $\K \not = 0$ the partition
function is zero, and the only way to get non-zero contribution to
the partition function is to either introduce the Dirac string, or
put the particle on a spatial boundary. Below we will argue that the
two are equivalent.

For similar reason the local symmetry breaking insertion, say at the
point $x$, also has to be connected by Dirac string to the boundary.
The easiest way to see this is to extract an integral over $dg(x)$
from the measure $\cD A$, where $g(x)$ is some holonomy connecting
$x$ to the reference point along some path. If $\cD A$ is the
Ashtekar-Lewandowski  measure \cite{AL}, it has to contain $dg(x)$.
For simplicity consider 1-point function. The argument however will
hold for n-point functions for non-coincident $x_n$ as well. We will
extract explicitly $g(x)$ variable, after which the 1-point function
can be written as
\begin{equation}\label{1point}
W(x)=\int \cD A dg(x) \tr\big( g(x)^{-1}\v g(x)\F(x)\wedge \F(x)
\big)\exp \Big(i\int F^{IJ}\wedge F_{IJ}\Big),
\end{equation}
where $\v$ is the matrix representation of the vector $v^A$, in an
appropriate representation. Now one can notice that $g(x)$ is a
functional of connection in the bulk, while the exponent term in the
absence of Dirac string does not depend on the bulk connection. This
means that the only dependence of the integrand in (\ref{1point}) on
$g$ is in pre-exponential term. The integral over $dg(x)$ can be
therefore explicitly calculated. It equals zero because it is an
averaging of a vector over all possible directions. The only way to
get a nonzero contribution is to introduce the Dirac string as in
the case of particle.

 The presence
of the Dirac strings makes the holonomies path-dependent: the
holonomy passing to the left from the string will be different from
the one passing to the right. To keep track of this information we
should to the list of all holonomies  relevant for the $\beta=0$
case described above add those around each string. A holonomy around
the string ending at $x_n$, $x_{p_i}$, or $x_{p_f}$ with the initial
point connected to the tip of the string will be called $g^*_n$,
$g^*_{p_i}$, and $g^*_{p_f}$ respectively. The natural choice of
pathes for holonomes $g_{n}$, $g_{p_i}$ and $g_{p_f}$ will be then
the following. $g_{n}$ for example will start at $x_n$ which is the
tip of one Dirac string, then follow this string towards the
boundary, and then go along the boundary to the reference point
$x_0$.

One can easily see that the above picture is in fact three
dimensional instead of four. A Dirac string (more precisely a thin
tube cut around it) can be considered as an extension of the
boundary of space. The tip of each Dirac string can be considered as
a puncture on the boundary. The holonomies along  the strings can be
then considered as holonomies taken along the boundary and
connecting punctures. And the holonomies around strings just become
holonomies around punctures. This dimensional reduction is, of
course, just a reflection of the fact that the action in the
exponent in (\ref{n-point1}) is the action of three dimensional
theory, namely the Chern-Simons theory.

In the present paper we will be interested only in the limit
$\alpha \to 0$. Quantum theory allows us to consistently take that
limit just by dropping all the terms proportional to $\alpha$ in
the expansion (\ref{series2}), no care about integrability
conditions as those encountered in classical theory is needed.

The relevant n-point (actually 0-point) function will simply read:

\begin{eqnarray}\label{n-point2}
W(\{x_{p_i}\},\{x_{p_f}\})=\int \cD A \prod\limits_p\big( {\cal
W}_p(x_{p_i},x_{p_f})\big) \exp [i \int_{\partial M} Y_{CS}(A)],
\end{eqnarray}
Following the discussion above in this expression we have taken into
account that all the points $x_{p_i}$ and $x_{p_f}$ and all the
Wilson lines connecting them can be mapped on the three dimensional
boundary of space.

Instead of doing path integral we will try to deduce physics by
making canonical analysis of the theory encoded in (\ref{n-point2}).
For this it is enough to consider the variable residing on the
initial slice of $\partial M$, ie those connecting $x_{p_i}$. Let
$x_{p_0}$ be a reference point, and $x_{p_1}$ is the particle of
interest. The position of the particle with respect to the reference
point will be described by an $SO(4,1)$ group element -- the
holonomy $g_{01}$ connecting $x_{p_0}$ and $x_{p_1}$. Notice that
although the expression for the amplitudes is described in terms of
three dimensional theory the particle still has 3 translational
degrees of freedom as does four dimensional particle.

Now one can ask what is the momentum of the particle, i.e. what is
the variable canonically conjugate to translational part of
$g_{01}$. The Poisson structure of Chern-Simons theory
\begin{equation}\label{pb}
\{A_i^{AB},A_j^{CD} \}=\beta \epsilon_{ij}
\delta^{[AC}\delta^{B]D},
\end{equation}
tells us that it has to be the translational part of the holonomy
$g^*_1$ which is taken around the particle $x_{p_1}$ which starts
and ends at $x_{p_0}$. $g^*_1$ is also an element of $SO(4,1)$
group, so the momentum space of the particle is de Sitter space.
This strongly indicates that a theory describing relativistic
symmetries of such a system of particles, in the flat space,
$\Lambda\rightarrow0$ limit will be of DSR type, like in the 2+1
dimensional case.

This expectation is also motivated by strong results on Hamiltonian
quantization of Chern-Simon theories with arbitrary gauge group on
Riemann surfaces \cite{ChSq}. On the basis of these results one
expects that after quantization the symmetry group $SO(4,1)$ will be
replaced by its quantum counter part $U_q(SO(4,1))$ (Note however
that Fock--Rosly theorem applies, strictly speaking, to compact
gauge group and its extension to the non-compact case is highly
nontrivial, see \cite{Buffenoir:2002tx} for the analysis of
$SL(2,C)$ case.) Then the limit of cosmological constant going to
zero will presumably lead to emergence of $\kappa$-Poincar\`e
algebra \cite{qP}, which is central for DSR program.

To finish this section let us say a few words about euclidian
version of this model. For $SO(5)$ gauge group the momentum space
is a sphere, both energy and momenta are bounded, and the theory
has a natural UV cutoff. Spacetime collapses at the energy higher
than $\sqrt{\Lambda}/\beta$. It is not yet clear what is the
precise mechanism of this collapse. There are two possibilities in
principle. Either all spacetime disappears at one instant, as it
is the case for $2+1$ gravity with zero cosmological constant. Or
it starts collapsing in some region around a particle, and the
region will be growing as the mass of the particle is growing. In
the later case there may be some effects not suppressed by the
smallness of the cosmological constant. This will be studied in
the subsequent papers.

\section{Discussion}

To this point we do not have explicit expressions for higher order
corrections to scattering amplitudes. In this section we will
briefly mention the problems one has to face in calculating them.

Let us try to reexpress (\ref{n-point1}) in terms of finite
dimensional integrals over holonomies instead of path integral
over connections. First  one can replace the action in the
exponent in (\ref{n-point1}) by Chern-Simons action. Then we can
take a locally flat connection connection $A(\{g\})$ whose moduli
space is defined by holonomies $g_n$, $g_{p_i}$, $g_{p_f}$ and
$g^*_n$, $g^*_{p_i}$,  $g^*_{p_f}$ (we will denote the whole set
by $\{ g\}$ ), and plug it into the Chern -Simons action. We will
have a functional depending on only finite number of parameters.

\begin{equation}
\int Y_{CS}(A(\{g\})) \equiv \bar S_{CS}(\{g\})
\end{equation}

A more complicated treatment is needed for preexponent terms. The
local curvature has somehow to be read of from the set of
holonomies. Clearly the two relevant holonomies at a point $x_n$ are
$g_n$ and $g^*_n$. They are dual to each other and therefore the
should give rise to two dual elements of curvature entering a local
$F_{BC}(x_n)\wedge F_{DE}(x_n)$ insertion in the action. The
difficulty here is that holonomies are group elements and curvature
is an algebra element. Therefore we need a certain map ${\cal
F}^{IJ}(g)$ from from Lie group to Lie algebra. Having such a map we
can substitute in our path integral
\begin{equation}\label{localins}
F_{BC}(x_n)\wedge F_{DE}(x_n) \to {\cal F}^{BC}(g_n){\cal
F}^{DE}(g^*_n)
\end{equation}
to obtain
\begin{eqnarray}\label{n-point11}
W(x_1,...,x_n,\{x_{p_i}\},\{x_{p_f}\})=\int \prod\limits{\{g\}}dg
\prod\limits_p\big( G_p(g_{p_i}^{-1}g_{p_f})\big)\nonumber
\\ \prod\limits_n \big( v_A \epsilon^{ABCDE}{\cal F}^{BC}(g_n){\cal
F}^{DE}(g^*_n)\big)  \exp [i \bar S_{CS}(\{g\})],
\end{eqnarray}
This is a ``Feynman diagram'' expression for an $n$-point function
in this perturbation theory. The only missing ingredient is the
exact form of the map ${\cal F}$. The only natural requirement on
this map is gauge invariance
\begin{equation}
{\cal F}^{IJ}(h^{-1}gh)=h_K^I{\cal F}^{KL}(g)h_L^J
\end{equation}
This however does not specify the map uniquely and it has to be
derived from other requirements. Our proposal would be to derive the
exact form of ${\cal F}$ by comparing an analog of the expression
(\ref{n-point11}) but without symmetry breaking with derivatives of
Crane -Yetter invariants with respect to its coupling constant.

\section*{Acknowledgements}
We thank Laurent Freidel for collaboration on a related topic. AS
thanks Renate Loll for discussions and Abhay Ashtekar for pointing
to some results relevant to this work. For JK-G this work  is
partially supported by the grant KBN 1 P03B 01828


\begin{thebibliography}{99.}

\bibitem{2+1grav} E.~Witten,
  ``(2+1)-dimensional gravity as an exactly soluble system,''
  Nucl.\ Phys.\ B {\bf 311} (1988) 46. S.~Carlip,
  ``Quantum gravity in 2+1 dimensions,'' Cambridge University Press,
  1998; S.~Carlip,
  ``Quantum gravity in 2+1 dimensions: The case of a closed universe,''
  Living Rev.\ Rel.\  {\bf 8} (2005) 1
  [arXiv:gr-qc/0409039]
\bibitem{giovanni}
 Giovanni Amelino-Camelia  , Lee Smolin, Artem
Starodubtsev  "Quantum symmetry, the cosmological constant and
Planck scale phenomenology."  Published in
Class.Quant.Grav.21:3095-3110,2004. e-Print Archive: hep-th/0306134,
Artem Starodubtsev, "Topological excitations around the vacuum of
quantum gravity. 1. The Symmetries of the vacuum." e-Print Archive:
hep-th/0306135; L.~Freidel, J.~Kowalski-Glikman and L.~Smolin,
  Phys.\ Rev.\ D {\bf 69}, 044001 (2004)
  [arXiv:hep-th/0307085].

\bibitem{dsr} G.~Amelino-Camelia, ``Testable scenario for relativity with minimum-length,''
Phys.\ Lett.\ B {\bf 510}, 255 (2001) [arXiv:hep-th/0012238]; for
review see J.~Kowalski-Glikman,
  ``Introduction to doubly special relativity,''
  Lect.\ Notes Phys.\  {\bf 669}, 131 (2005)
  [arXiv:hep-th/0405273] and J.~Kowalski-Glikman,
  ``Doubly special relativity: Facts and prospects,''
  arXiv:gr-qc/0603022

\bibitem{Freidel:2005ak}
  L.~Freidel and A.~Starodubtsev,
  ``Quantum gravity in terms of topological observables,''
  arXiv:hep-th/0501191.

\bibitem{Freidel:2006hv}
  L.~Freidel, J.~Kowalski-Glikman and A.~Starodubtsev,
  ``Particles as Wilson lines of gravitational field,''
  arXiv:gr-qc/0607014.

\bibitem{Smolin:2003qu}
  L.~Smolin and A.~Starodubtsev,
  ``General relativity with a topological phase: An action principle,''
  arXiv:hep-th/0311163.

\bibitem{ksx}
G.C. Debney, R.P. Kerr, and A. Schild, J. Math. Phys., 10, 1842
(1969), Basilis C. Xanthopolous, "Exact vacuum solutions of
Einstein's equations from linearized solutions", J. Math. Phys.,
19, 1607 (1978)


\bibitem{rovelli1}
 Carlo Rovelli , Simone Speziale,  "On the perturbative expansion of a quantum field theory around a
topological sector" e-Print Archive: gr-qc/0508106


\bibitem{bfstrings}John C.
Baez , Alejandro Perez  "Quantization of strings and branes
coupled to BF theory"  e-Print Archive: gr-qc/0605087


\bibitem{nakahara} M.\ Nakahara, {\em Geometry, Topology, and Physics}, II ed., IoP Publishing, Bristol, 2005.


\bibitem{rovelli2}
Carlo Rovelli, "Graviton propagator from background-independent
quantum gravity."  Phys.Rev.Lett.97:151301,2006. e-Print Archive:
gr-qc/0508124

\bibitem{AL}
Abhay Ashtekar, Jerzy Lewandowski, "Projective techniques and
functional integration for gauge theories."
J.Math.Phys.36:2170-2191,1995. e-Print Archive: gr-qc/9411046

Jerzy Lewandowski, Andrzej Okolow, Hanno Sahlmann, Thomas Thiemann
"Uniqueness of diffeomorphism invariant states on holonomy-flux
algebras."  e-Print Archive: gr-qc/0504147

\bibitem{ChSq} V.~V.~Fock and A.~A.~Rosly,
  ``Poisson structure on moduli of flat connections on Riemann surfaces and
  r-matrix,''
  Am.\ Math.\ Soc.\ Transl.\  {\bf 191} (1999) 67
  [arXiv:math.qa/9802054]; A.~Y.~Alekseev, H.~Grosse and V.~Schomerus,
  ``Combinatorial quantization of the Hamiltonian Chern-Simons theory,''
  Commun.\ Math.\ Phys.\  {\bf 172}, 317 (1995)
  [arXiv:hep-th/9403066], A.~Y.~Alekseev, H.~Grosse and V.~Schomerus,
  ``Combinatorial quantization of the Hamiltonian Chern-Simons theory. 2,''
  Commun.\ Math.\ Phys.\  {\bf 174}, 561 (1995)
  [arXiv:hep-th/9408097]

  \bibitem{Buffenoir:2002tx}
  E.~Buffenoir, K.~Noui and P.~Roche,
  ``Hamiltonian quantization of Chern-Simons theory with SL(2,C) group,''
  Class.\ Quant.\ Grav.\  {\bf 19}, 4953 (2002)
  [arXiv:hep-th/0202121].

\bibitem{qP} J.~Lukierski, H.~Ruegg, A.~Nowicki and V.~N.~Tolstoi,
``Q deformation of Poincar\'e algebra,'' Phys.\ Lett.\ B {\bf 264}
(1991) 331; J.~Lukierski, A.~Nowicki and H.~Ruegg, ``New quantum
Poincare algebra and k deformed field theory,'' Phys.\ Lett.\ B {\bf
293} (1992) 344



\end{thebibliography}
\end{document}